\documentclass[aps,prb,superscriptaddress,twocolumn,showpacs]{revtex4}
\usepackage{amsmath}
\usepackage{amssymb}
\usepackage{dsfont}
\usepackage{graphicx}
\usepackage{subfigure}
\usepackage{tikz}
\tikzset{>=stealth}
\usetikzlibrary{arrows,matrix,decorations.markings}

\newcommand{\Tmv}[2]{T_{#1 \rightarrow #2}}

\newcommand{\Tact}[1]{\mathcal{T}\Biggl(\begin{matrix}#1\end{matrix}\Biggr)}
\newcommand{\rmu}{\mathrm{u}}

\newcommand{\arl}{\ar@{-}|@{>}}
\newcommand{\arr}{\ar@{-}|@{<}}
\newcommand{\aline}{\ar@{-}}
\newcommand{\rmd}{\mathrm{d}}

\newcommand{\rmv}{\mathrm{v}}

\newcommand{\Qb}{\overline{Q}}
\newcommand{\Bb}{\overline{B}}

\newcommand{\ket}[1]{\left|{#1}\right\rangle}

\newcommand{\bket}[1]{\Biggl|{\bmm #1 \emm}\Biggr\rangle}

\newcommand{\bpm}{\begin{pmatrix}}
\newcommand{\epm}{\end{pmatrix}}
\newcommand{\bmm}{\begin{matrix}}
\newcommand{\emm}{\end{matrix}}

\usetikzlibrary{decorations.markings}
\tikzset{->-/.style={decoration={
			markings,
			mark=at position .55 with {\arrow{stealth}}},postaction={decorate}}}

\usetikzlibrary{decorations.markings}
\tikzset{->-/.style={decoration={
			markings,
			mark=at position .55 with {\arrow{stealth}}},postaction={decorate}}}

\newcommand{\ExpBimoduleDef}{
	\Biggl|\;\begin{matrix}
		\begin{tikzpicture}[scale=0.8]
		\draw [->-] (1.6,9.8) -- (2.4,9.6);
		\node [] at (2.06,9.93) {\scriptsize $b_{2}$};
		\draw [->-] (2.4,8.) -- (2.4,8.8);
		\node [] at (2.11,8.4) {\scriptsize $j_{1}$};
		\draw [->-] (2.4,9.6) -- (2.4,10.4);
		\node [] at (2.69,10.) {\scriptsize $j_{2}$};
		\draw [->-] (3.2,8.6) -- (2.4,8.8);
		\node [] at (2.74,8.47) {\scriptsize $a_{1}$};
		\draw [->-] (3.2,9.4) -- (2.4,9.6);
		\node [] at (2.74,9.27) {\scriptsize $a_{2}$};
		\draw [very thick] (2.4,8.8) -- (2.4,9.6);
		\draw [->-] (1.6,9.) -- (2.4,8.8);
		\node [] at (2.06,9.13) {\scriptsize $b_{1}$};
		\node [fill=white,draw,rounded corners=.06cm] at (2.4,8.8) {\scriptsize $P$};
		\node [fill=white,draw,rounded corners=.06cm] at (2.4,9.6) {\scriptsize $P$};
		\end{tikzpicture}\:
	\end{matrix}\Biggr\rangle
}
\newcommand{\ExpBimoduleDefAA}{
	\Biggl|\;\begin{matrix}
		\begin{tikzpicture}[scale=0.8]
		\draw [->-] (2.8,8.) -- (2.8,9.2);
		\node [] at (2.51,8.6) {\scriptsize $j_{1}$};
		\draw [->-] (2.8,9.2) -- (2.8,10.4);
		\node [] at (3.09,9.8) {\scriptsize $j_{2}$};
		\draw [->-] (3.5,8.2) -- (3.4,8.8);
		\node [] at (3.16,8.45) {\scriptsize $a_{1}$};
		\draw [->-] (4.,8.6) -- (3.4,8.8);
		\node [] at (3.78,8.93) {\scriptsize $a_{2}$};
		\draw [very thick] (2.2,9.6) -- (2.8,9.2);
		\draw [very thick] (3.4,8.8) -- (2.8,9.2);
		\draw [->-] (1.6,9.8) -- (2.2,9.6);
		\node [] at (1.82,9.47) {\scriptsize $b_{1}$};
		\draw [->-] (2.,10.2) -- (2.2,9.6);
		\node [] at (2.37,9.99) {\scriptsize $b_{2}$};
		\draw [fill] (3.4,8.8) circle [x radius=0.08, y radius=0.08];
		\draw [fill] (2.2,9.6) circle [x radius=0.08, y radius=0.08];
		\node [fill=white,draw,rounded corners=.06cm] at (2.8,9.2) {\scriptsize $P$};
		\end{tikzpicture}\:
	\end{matrix}\Biggr\rangle
}
\newcommand{\ExpBoundaryBbpAA}{
	\Biggl|\;\begin{matrix}
		\begin{tikzpicture}[scale=0.7]
		\path [lightgray, fill] (1.6,7.2) -- (2.5,7.2) -- (2.9,8.) -- (2.5,8.8) -- (2.9,9.6) -- (2.5,10.4) -- (1.6,10.4) -- cycle;
		\draw [->-] (1.9,8.8) -- (2.5,8.8);
		\node [] at (2.2,8.58) {\scriptsize $j_{5}$};
		\draw [->-] (2.5,8.8) -- (2.9,9.6);
		\node [] at (2.44,9.33) {\scriptsize $j_{3}$};
		\draw [->-] (2.9,8.) -- (2.5,8.8);
		\node [] at (2.96,8.53) {\scriptsize $j_{2}$};
		\draw [->-] (2.9,8.) -- (3.7,8.);
		\node [] at (3.3,7.78) {\scriptsize $a_{1}$};
		\draw [->-] (2.9,9.6) -- (2.5,10.4);
		\node [] at (2.96,10.13) {\scriptsize $j_{4}$};
		\draw [->-] (2.9,9.6) -- (3.7,9.6);
		\node [] at (3.3,9.38) {\scriptsize $a_{2}$};
		\draw [->-] (2.5,7.2) -- (2.9,8.);
		\node [] at (2.44,7.73) {\scriptsize $j_{1}$};
		\end{tikzpicture}\:
	\end{matrix}\Biggr\rangle
}
\newcommand{\BoundaryBbpAF}{
	\begin{tikzpicture}[scale=0.7]
	\path [lightgray, fill] (1.6,7.2) -- (2.5,7.2) -- (2.9,8.) -- (2.5,8.8) -- (2.9,9.6) -- (2.9,9.6) -- (2.5,10.4) -- (1.6,10.4) -- cycle;
	\draw [->-] (1.9,8.8) -- (2.5,8.8);
	\node [] at (2.2,8.58) {\scriptsize $j_{5}$};
	\draw [->-] (2.5,8.8) -- (2.9,9.6);
	\node [] at (2.39,9.36) {\scriptsize $j'_{3}$};
	\draw [->-] (2.9,9.6) -- (2.5,10.4);
	\node [] at (2.96,10.13) {\scriptsize $j_{4}$};
	\draw [->-] (2.9,9.6) -- (3.5,9.6);
	\node [] at (3.2,9.38) {\scriptsize $a'_{2}$};
	\draw [->-] (2.5,7.2) -- (2.9,8.);
	\node [] at (2.44,7.73) {\scriptsize $j_{1}$};
	\draw [->-] (2.9,8.) -- (2.5,8.8);
	\node [] at (3.01,8.56) {\scriptsize $j'_{2}$};
	\draw [->-] (2.9,8.) -- (3.5,8.);
	\node [] at (3.2,7.78) {\scriptsize $a'_{1}$};
	\end{tikzpicture}
}

\newcommand{\ExpBoundaryBbpAG}{
	\sum_{ta'_{1}a'_{2}}\frac{\sqrt{D}\mathrm{u}_{a'_{1}}\mathrm{u}_{a'_{2}}}{\mathrm{d}_A\mathrm{u}_{a_{1}}\mathrm{u}_{a_{2}}}
	\Biggl|\;\begin{matrix}
		\begin{tikzpicture}[scale=0.7]
		\path [lightgray, fill] (1.8,7.2) -- (2.5,7.2) -- (2.9,8.) -- (2.5,8.8) -- (2.9,9.6) -- (2.5,10.4) -- (1.8,10.4) -- cycle;
		\draw [->-] (2.,8.8) -- (2.5,8.8);
		\node [] at (2.25,8.58) {\scriptsize $j_{5}$};
		\draw [->-] (2.5,7.2) -- (2.9,8.);
		\node [] at (2.44,7.73) {\scriptsize $j_{1}$};
		\draw [->-] (2.5,8.8) -- (2.9,9.6);
		\node [] at (2.44,9.33) {\scriptsize $j_{3}$};
		\draw [->-] (2.9,8.) -- (2.5,8.8);
		\node [] at (2.96,8.53) {\scriptsize $j_{2}$};
		\draw [->-] (2.9,8.) -- (3.7,8.);
		\node [] at (3.3,7.78) {\scriptsize $a_{1}$};
		\draw [->-] (2.9,9.6) -- (2.5,10.4);
		\node [] at (2.44,9.87) {\scriptsize $j_{4}$};
		\draw [->-] (2.9,9.6) -- (3.7,9.6);
		\node [] at (3.3,9.82) {\scriptsize $a_{2}$};
		\draw [->-] (3.7,8.) -- (4.5,8.);
		\node [] at (4.1,7.78) {\scriptsize $a'_{1}$};
		\draw [->-] (3.7,9.6) -- (4.5,9.6);
		\node [] at (4.1,9.82) {\scriptsize $a'_{2}$};
		\draw [->-] (3.7,9.6) -- (3.7,8.);
		\node [] at (3.92,8.8) {\scriptsize $t$};
		\draw [fill] (3.7,9.6) circle [x radius=0.08, y radius=0.08];
		\draw [fill] (3.7,8.) circle [x radius=0.08, y radius=0.08];
		\node [] at (3.3,8.9) {\scriptsize $\times$};
		\end{tikzpicture}\:
	\end{matrix}\Biggr\rangle
}
\newcommand{\FigBoundary}{
	\begin{tikzpicture}[scale=0.65]
	\path [lightgray, fill] (1.6,4.8) -- (3.8,4.8) -- (3.4,5.6) -- (3.8,6.4) -- (3.4,7.2) -- (3.8,8.) -- (3.4,8.8) -- (3.8,9.6) -- (3.4,10.4) -- (1.6,10.4) -- (1.6,8.9) -- cycle;
	\draw [->-] (2.6,5.6) -- (3.4,5.6);
	\node [] at (3.,5.38) {\scriptsize $j_{2}$};
	\draw [->-] (2.6,7.2) -- (3.4,7.2);
	\node [] at (3.,6.98) {\scriptsize $j_{5}$};
	\draw [->-] (2.6,8.8) -- (3.4,8.8);
	\node [] at (3.,8.58) {\scriptsize $j_{8}$};
	\draw [->-] (2.6,10.4) -- (3.4,10.4);
	\node [] at (3.,10.18) {\scriptsize $j_{11}$};
	\draw [->-] (3.4,5.6) -- (3.8,6.4);
	\node [] at (3.34,6.13) {\scriptsize $j_{3}$};
	\draw [->-] (3.4,7.2) -- (3.8,8.);
	\node [] at (3.34,7.73) {\scriptsize $j_{6}$};
	\draw [->-] (3.8,4.8) -- (3.4,5.6);
	\node [] at (3.86,5.33) {\scriptsize $j_{1}$};
	\draw [->-] (3.8,6.4) -- (3.4,7.2);
	\node [] at (3.86,6.93) {\scriptsize $j_{4}$};
	\draw [->-] (3.8,6.4) -- (4.6,6.4);
	\node [] at (4.2,6.18) {\scriptsize $a_{1}$};
	\draw [->-] (3.8,8.) -- (3.4,8.8);
	\node [] at (3.86,8.53) {\scriptsize $j_{7}$};
	\draw [->-] (3.8,8.) -- (4.6,8.);
	\node [] at (4.2,7.78) {\scriptsize $a_{2}$};
	\draw [->-] (3.8,9.6) -- (3.4,10.4);
	\node [] at (3.91,10.16) {\scriptsize $j_{10}$};
	\draw [->-] (3.8,9.6) -- (4.6,9.6);
	\node [] at (4.2,9.38) {\scriptsize $a_{3}$};
	\draw [->-] (3.4,8.8) -- (3.8,9.6);
	\node [] at (3.34,9.33) {\scriptsize $j_{9}$};
	\node [] at (2.,9.2) {\scriptsize $\text{bulk}$};
	\end{tikzpicture}
}

\newcommand{\FigBoundaryModuleAA}{
	\begin{tikzpicture}[scale=1]
	\draw [lightgray, fill] (3.2,7.2) ..controls (3.481,7.187) and (3.608,7.197) .. (3.8,7.2)
	..controls (3.992,7.203) and (4.248,7.2) .. (4.5,7.2)
	..controls (4.752,7.2) and (5.002,7.203) .. (5.2,7.2)
	..controls (5.398,7.197) and (5.546,7.187) .. (5.7,7.2)
	..controls (5.854,7.213) and (6.013,7.248) .. (6.2,7.3)
	..controls (6.387,7.352) and (6.601,7.421) .. (6.7,7.6)
	..controls (6.799,7.779) and (6.781,8.067) .. (6.8,8.3)
	..controls (6.819,8.533) and (6.874,8.711) .. (6.9,8.9)
	..controls (6.926,9.089) and (6.923,9.291) .. (6.9,9.5)
	..controls (6.877,9.709) and (6.834,9.926) .. (6.7,10.1)
	..controls (6.566,10.274) and (6.34,10.403) .. (6.2,10.4)
	..controls (6.06,10.397) and (6.007,10.26) .. (5.9,10.2)
	..controls (5.793,10.14) and (5.633,10.156) .. (5.5,10.2)
	..controls (5.367,10.244) and (5.26,10.315) .. (5.1,10.3)
	..controls (4.94,10.285) and (4.726,10.184) .. (4.5,10.2)
	..controls (4.274,10.216) and (4.035,10.35) .. (3.8,10.4)
	..controls (3.565,10.45) and (3.333,10.415) .. (3.,10.4)
	..controls (2.667,10.385) and (2.232,10.389) .. (2.,10.2)
	..controls (1.768,10.011) and (1.739,9.629) .. (1.7,9.3)
	..controls (1.661,8.971) and (1.611,8.696) .. (1.6,8.4)
	..controls (1.589,8.104) and (1.618,7.789) .. (1.7,7.6)
	..controls (1.782,7.411) and (1.916,7.349) .. (2.2,7.3)
	..controls (2.484,7.251) and (2.919,7.213) .. (3.2,7.2);
	\draw [white, fill] (2.4,8.9) ..controls (2.264,8.896) and (2.178,8.8) .. (2.1,8.7)
	..controls (2.022,8.6) and (1.954,8.494) .. (1.9,8.4)
	..controls (1.846,8.306) and (1.806,8.222) .. (1.8,8.1)
	..controls (1.794,7.978) and (1.821,7.816) .. (2.,7.8)
	..controls (2.179,7.784) and (2.51,7.914) .. (2.7,8.)
	..controls (2.89,8.086) and (2.941,8.127) .. (3.,8.2)
	..controls (3.059,8.273) and (3.128,8.378) .. (3.2,8.5)
	..controls (3.272,8.622) and (3.348,8.759) .. (3.3,8.8)
	..controls (3.252,8.841) and (3.079,8.784) .. (2.9,8.8)
	..controls (2.721,8.816) and (2.536,8.904) .. (2.4,8.9);
	\draw [white, fill] (2.9,9.9) ..controls (2.793,9.831) and (2.693,9.786) .. (2.6,9.7)
	..controls (2.507,9.614) and (2.42,9.487) .. (2.4,9.4)
	..controls (2.38,9.313) and (2.426,9.267) .. (2.5,9.2)
	..controls (2.574,9.133) and (2.675,9.044) .. (2.8,9.)
	..controls (2.925,8.956) and (3.075,8.958) .. (3.2,9.)
	..controls (3.325,9.042) and (3.425,9.123) .. (3.5,9.2)
	..controls (3.575,9.277) and (3.625,9.35) .. (3.7,9.5)
	..controls (3.775,9.65) and (3.874,9.878) .. (3.8,10.)
	..controls (3.726,10.122) and (3.48,10.138) .. (3.3,10.1)
	..controls (3.12,10.062) and (3.007,9.969) .. (2.9,9.9);
	\draw [white, fill] (4.9,9.3) ..controls (4.784,9.246) and (4.733,9.151) .. (4.7,9.1)
	..controls (4.667,9.049) and (4.654,9.041) .. (4.6,9.)
	..controls (4.546,8.959) and (4.451,8.884) .. (4.5,8.8)
	..controls (4.549,8.716) and (4.741,8.622) .. (4.9,8.6)
	..controls (5.059,8.578) and (5.184,8.628) .. (5.3,8.7)
	..controls (5.416,8.772) and (5.522,8.867) .. (5.6,8.9)
	..controls (5.678,8.933) and (5.728,8.904) .. (5.8,9.)
	..controls (5.872,9.096) and (5.967,9.316) .. (5.9,9.4)
	..controls (5.833,9.484) and (5.604,9.433) .. (5.4,9.4)
	..controls (5.196,9.367) and (5.016,9.354) .. (4.9,9.3);
	\draw [white, fill] (4.,8.1) ..controls (3.85,8.051) and (3.703,7.994) .. (3.8,8.)
	..controls (3.897,8.006) and (4.24,8.076) .. (4.2,8.)
	..controls (4.16,7.924) and (3.737,7.704) .. (3.7,7.6)
	..controls (3.663,7.496) and (4.011,7.508) .. (4.2,7.5)
	..controls (4.389,7.492) and (4.418,7.463) .. (4.5,7.5)
	..controls (4.582,7.537) and (4.716,7.638) .. (4.9,7.7)
	..controls (5.084,7.762) and (5.318,7.785) .. (5.5,7.8)
	..controls (5.682,7.815) and (5.813,7.824) .. (6.,7.9)
	..controls (6.187,7.976) and (6.43,8.12) .. (6.5,8.3)
	..controls (6.57,8.48) and (6.466,8.695) .. (6.4,8.9)
	..controls (6.334,9.105) and (6.306,9.301) .. (6.2,9.3)
	..controls (6.094,9.299) and (5.91,9.101) .. (5.9,8.9)
	..controls (5.89,8.699) and (6.055,8.494) .. (6.,8.4)
	..controls (5.945,8.306) and (5.672,8.322) .. (5.4,8.3)
	..controls (5.128,8.278) and (4.857,8.217) .. (4.7,8.2)
	..controls (4.543,8.183) and (4.498,8.208) .. (4.4,8.2)
	..controls (4.302,8.192) and (4.15,8.149) .. (4.,8.1);
	\draw [thick] (2.4,8.9) ..controls (2.264,8.896) and (2.178,8.8) .. (2.1,8.7)
	..controls (2.022,8.6) and (1.954,8.494) .. (1.9,8.4)
	..controls (1.846,8.306) and (1.806,8.222) .. (1.8,8.1)
	..controls (1.794,7.978) and (1.821,7.816) .. (2.,7.8)
	..controls (2.179,7.784) and (2.51,7.914) .. (2.7,8.)
	..controls (2.89,8.086) and (2.941,8.127) .. (3.,8.2)
	..controls (3.059,8.273) and (3.128,8.378) .. (3.2,8.5)
	..controls (3.272,8.622) and (3.348,8.759) .. (3.3,8.8)
	..controls (3.252,8.841) and (3.079,8.784) .. (2.9,8.8)
	..controls (2.721,8.816) and (2.536,8.904) .. (2.4,8.9);
	\draw [thick] (2.9,9.9) ..controls (2.793,9.831) and (2.693,9.786) .. (2.6,9.7)
	..controls (2.507,9.614) and (2.42,9.487) .. (2.4,9.4)
	..controls (2.38,9.313) and (2.426,9.267) .. (2.5,9.2)
	..controls (2.574,9.133) and (2.675,9.044) .. (2.8,9.)
	..controls (2.925,8.956) and (3.075,8.958) .. (3.2,9.)
	..controls (3.325,9.042) and (3.425,9.123) .. (3.5,9.2)
	..controls (3.575,9.277) and (3.625,9.35) .. (3.7,9.5)
	..controls (3.775,9.65) and (3.874,9.878) .. (3.8,10.)
	..controls (3.726,10.122) and (3.48,10.138) .. (3.3,10.1)
	..controls (3.12,10.062) and (3.007,9.969) .. (2.9,9.9);
	\draw [thick] (4.9,9.3) ..controls (4.784,9.246) and (4.733,9.151) .. (4.7,9.1)
	..controls (4.667,9.049) and (4.654,9.041) .. (4.6,9.)
	..controls (4.546,8.959) and (4.451,8.884) .. (4.5,8.8)
	..controls (4.549,8.716) and (4.741,8.622) .. (4.9,8.6)
	..controls (5.059,8.578) and (5.184,8.628) .. (5.3,8.7)
	..controls (5.416,8.772) and (5.522,8.867) .. (5.6,8.9)
	..controls (5.678,8.933) and (5.728,8.904) .. (5.8,9.)
	..controls (5.872,9.096) and (5.967,9.316) .. (5.9,9.4)
	..controls (5.833,9.484) and (5.604,9.433) .. (5.4,9.4)
	..controls (5.196,9.367) and (5.016,9.354) .. (4.9,9.3);
	\draw [thick] (4.,8.1) ..controls (3.853,8.051) and (3.703,7.994) .. (3.8,8.)
	..controls (3.897,8.006) and (4.24,8.076) .. (4.2,8.)
	..controls (4.16,7.924) and (3.737,7.704) .. (3.7,7.6)
	..controls (3.663,7.496) and (4.011,7.508) .. (4.2,7.5)
	..controls (4.389,7.492) and (4.418,7.463) .. (4.5,7.5)
	..controls (4.582,7.537) and (4.716,7.638) .. (4.9,7.7)
	..controls (5.084,7.762) and (5.318,7.785) .. (5.5,7.8)
	..controls (5.682,7.815) and (5.813,7.824) .. (6.,7.9)
	..controls (6.187,7.976) and (6.43,8.12) .. (6.5,8.3)
	..controls (6.57,8.48) and (6.465,8.694) .. (6.4,8.9)
	..controls (6.335,9.106) and (6.308,9.302) .. (6.2,9.3)
	..controls (6.092,9.298) and (5.903,9.099) .. (5.9,8.9)
	..controls (5.897,8.701) and (6.079,8.502) .. (6.,8.4)
	..controls (5.921,8.298) and (5.58,8.294) .. (5.4,8.3)
	..controls (5.22,8.306) and (5.199,8.322) .. (5.1,8.3)
	..controls (5.001,8.278) and (4.823,8.217) .. (4.7,8.2)
	..controls (4.577,8.183) and (4.507,8.208) .. (4.4,8.2)
	..controls (4.293,8.192) and (4.147,8.149) .. (4.,8.1);
	\draw [thick] (3.2,7.2) ..controls (3.481,7.187) and (3.608,7.197) .. (3.8,7.2)
	..controls (3.992,7.203) and (4.248,7.2) .. (4.5,7.2)
	..controls (4.752,7.2) and (5.002,7.203) .. (5.2,7.2)
	..controls (5.398,7.197) and (5.546,7.187) .. (5.7,7.2)
	..controls (5.854,7.213) and (6.013,7.248) .. (6.2,7.3)
	..controls (6.387,7.352) and (6.601,7.421) .. (6.7,7.6)
	..controls (6.799,7.779) and (6.781,8.067) .. (6.8,8.3)
	..controls (6.819,8.533) and (6.874,8.711) .. (6.9,8.9)
	..controls (6.926,9.089) and (6.923,9.291) .. (6.9,9.5)
	..controls (6.877,9.709) and (6.834,9.926) .. (6.7,10.1)
	..controls (6.566,10.274) and (6.34,10.403) .. (6.2,10.4)
	..controls (6.06,10.397) and (6.007,10.26) .. (5.9,10.2)
	..controls (5.793,10.14) and (5.633,10.156) .. (5.5,10.2)
	..controls (5.367,10.244) and (5.26,10.315) .. (5.1,10.3)
	..controls (4.94,10.285) and (4.726,10.184) .. (4.5,10.2)
	..controls (4.274,10.216) and (4.035,10.35) .. (3.8,10.4)
	..controls (3.565,10.45) and (3.333,10.415) .. (3.,10.4)
	..controls (2.667,10.385) and (2.232,10.389) .. (2.,10.2)
	..controls (1.768,10.011) and (1.739,9.629) .. (1.7,9.3)
	..controls (1.661,8.971) and (1.611,8.696) .. (1.6,8.4)
	..controls (1.589,8.104) and (1.618,7.789) .. (1.7,7.6)
	..controls (1.782,7.411) and (1.916,7.349) .. (2.2,7.3)
	..controls (2.484,7.251) and (2.919,7.213) .. (3.2,7.2);
	\node [] at (2.4,8.4) {\scriptsize $Mod_A$};
	\node [] at (3.1,9.5) {\scriptsize $Mod_A$};
	\node [] at (5.,8.9) {\scriptsize $Mod_B$};
	\node [] at (4.5,7.8) {\scriptsize $Mod_C$};
	\node [] at (4.4,10.4) {\scriptsize $Mod_D$};
	\end{tikzpicture}
}

\newcommand{\FigBoundaryTopoInvarianceAA}{
	\begin{tikzpicture}[scale=1]
	\path [lightgray, fill] (3.6,6.8) -- (4.1,6.8) -- (4.,7.1) -- (4.1,7.4) -- (4.,7.7) -- (4.1,8.) -- (4.,8.3) -- (4.1,8.6) -- (4.,8.9) -- (4.1,9.2) -- (4.,9.5) -- (4.1,9.8) -- (4.,10.1) -- (4.1,10.4) -- (3.6,10.4) -- cycle;
	\path [lightgray, fill] (1.6,6.8) -- (2.1,6.8) -- (2.,7.1) -- (2.1,7.4) -- (2.,7.7) -- (2.1,8.) -- (2.,8.3) -- (2.1,8.6) -- (2.,8.9) -- (2.1,9.2) -- (2.,9.5) -- (2.1,9.8) -- (2.,10.1) -- (2.1,10.4) -- (1.6,10.4) -- cycle;
	\draw [] (5.5,9.8) ..controls (5.522,9.64) and (5.544,9.48) .. (5.6,9.4)
	..controls (5.656,9.32) and (5.744,9.32) .. (5.8,9.4)
	..controls (5.856,9.48) and (5.878,9.64) .. (5.9,9.8);
	\draw [] (5.5,9.8) ..controls (5.522,9.96) and (5.544,10.12) .. (5.6,10.2)
	..controls (5.656,10.28) and (5.744,10.28) .. (5.8,10.2)
	..controls (5.856,10.12) and (5.878,9.96) .. (5.9,9.8);
	\draw [->-] (4.1,7.4) ..controls (4.325,7.392) and (4.55,7.383) .. (4.7,7.4)
	..controls (4.85,7.417) and (4.925,7.458) .. (5.,7.5);
	\draw [->-] (4.1,8.) ..controls (4.233,7.983) and (4.367,7.967) .. (4.5,8.)
	..controls (4.633,8.033) and (4.767,8.117) .. (4.9,8.2);
	\draw [->-] (4.1,8.6) ..controls (4.233,8.583) and (4.367,8.567) .. (4.5,8.5)
	..controls (4.633,8.433) and (4.767,8.317) .. (4.9,8.2);
	\draw [->-] (4.1,9.2) ..controls (4.192,9.183) and (4.283,9.167) .. (4.4,9.2)
	..controls (4.517,9.233) and (4.658,9.317) .. (4.8,9.4);
	\draw [->-] (4.1,9.8) ..controls (4.192,9.783) and (4.283,9.767) .. (4.4,9.7)
	..controls (4.517,9.633) and (4.658,9.517) .. (4.8,9.4);
	\draw [] (2.,7.1) -- (2.1,7.4);
	\draw [] (2.,7.7) -- (2.1,8.);
	\draw [] (2.,8.3) -- (2.1,8.6);
	\draw [] (2.,8.9) -- (2.1,9.2);
	\draw [] (2.,9.5) -- (2.1,9.8);
	\draw [] (2.,10.1) -- (2.1,10.4);
	\draw [] (2.1,6.8) -- (2.,7.1);
	\draw [] (2.1,7.4) -- (2.,7.7);
	\draw [] (2.1,8.) -- (2.,8.3);
	\draw [] (2.1,8.6) -- (2.,8.9);
	\draw [] (2.1,9.2) -- (2.,9.5);
	\draw [] (2.1,9.8) -- (2.,10.1);
	\draw [] (4.,7.1) -- (4.1,7.4);
	\draw [] (4.,7.7) -- (4.1,8.);
	\draw [] (4.,8.3) -- (4.1,8.6);
	\draw [] (4.,8.9) -- (4.1,9.2);
	\draw [] (4.,9.5) -- (4.1,9.8);
	\draw [] (4.,10.1) -- (4.1,10.4);
	\draw [] (4.1,6.8) -- (4.,7.1);
	\draw [] (4.1,7.4) -- (4.,7.7);
	\draw [] (4.1,8.) -- (4.,8.3);
	\draw [] (4.1,8.6) -- (4.,8.9);
	\draw [] (4.1,9.2) -- (4.,9.5);
	\draw [] (4.1,9.8) -- (4.,10.1);
	\draw [] (4.8,9.4) -- (4.9,9.);
	\draw [] (4.9,8.2) -- (5.3,8.2);
	\draw [] (4.9,9.) -- (5.3,8.7);
	\draw [] (4.9,9.) -- (5.4,9.2);
	\draw [] (5.,7.5) -- (5.3,8.2);
	\draw [] (5.,7.5) -- (5.6,7.4);
	\draw [] (5.3,8.2) -- (5.3,8.7);
	\draw [] (5.3,8.7) -- (5.7,8.5);
	\draw [] (5.4,9.2) -- (5.5,9.8);
	\draw [] (5.4,9.2) -- (5.7,8.9);
	\draw [] (5.6,7.4) -- (5.8,7.8);
	\draw [] (5.6,7.4) -- (6.,7.4);
	\draw [] (5.7,8.5) -- (5.7,8.9);
	\draw [] (5.7,8.5) -- (6.,8.2);
	\draw [] (5.8,7.8) -- (6.,8.2);
	\draw [] (5.8,7.8) -- (6.,7.4);
	\draw [] (5.7,8.9) -- (6.1,9.);
	\draw [] (6.,8.2) -- (6.7,8.2);
	\draw [] (5.9,9.8) -- (6.2,9.9);
	\draw [] (6.1,9.) -- (6.2,9.9);
	\draw [] (6.1,9.) -- (6.7,9.);
	\draw [] (6.,7.4) -- (6.7,7.4);
	\draw [] (6.2,9.9) -- (6.7,9.9);
	\draw [->-] (2.1,7.4) -- (2.5,7.4);
	\draw [->-] (2.1,8.) -- (2.5,8.);
	\draw [->-] (2.1,8.6) -- (2.5,8.6);
	\draw [->-] (2.1,9.2) -- (2.5,9.2);
	\draw [->-] (2.1,9.8) -- (2.5,9.8);
	\draw [gray, dashed, very thick] (4.6,6.8) -- (6.4,6.8);
	\draw [gray, dashed, very thick] (4.6,10.4) -- (4.6,6.8);
	\draw [gray, dashed, very thick] (6.4,6.8) -- (6.4,10.4);
	\draw [gray, dashed, very thick] (6.4,10.4) -- (4.6,10.4);
	\draw [very thick, ->] (2.8,8.7) -- (3.3,8.7);
	\draw [fill] (4.8,9.4) circle [x radius=0.08, y radius=0.08];
	\draw [fill] (4.9,8.2) circle [x radius=0.08, y radius=0.08];
	\draw [fill] (4.9,9.) circle [x radius=0.08, y radius=0.08];
	\draw [fill] (5.,7.5) circle [x radius=0.08, y radius=0.08];
	\draw [fill] (5.3,8.2) circle [x radius=0.08, y radius=0.08];
	\draw [fill] (5.3,8.7) circle [x radius=0.08, y radius=0.08];
	\draw [fill] (5.4,9.2) circle [x radius=0.08, y radius=0.08];
	\draw [fill] (5.6,7.4) circle [x radius=0.08, y radius=0.08];
	\draw [fill] (5.5,9.8) circle [x radius=0.08, y radius=0.08];
	\draw [fill] (5.7,8.5) circle [x radius=0.08, y radius=0.08];
	\draw [fill] (5.8,7.8) circle [x radius=0.08, y radius=0.08];
	\draw [fill] (5.7,8.9) circle [x radius=0.08, y radius=0.08];
	\draw [fill] (6.,8.2) circle [x radius=0.08, y radius=0.08];
	\draw [fill] (5.9,9.8) circle [x radius=0.08, y radius=0.08];
	\draw [fill] (6.1,9.) circle [x radius=0.08, y radius=0.08];
	\draw [fill] (6.,7.4) circle [x radius=0.08, y radius=0.08];
	\draw [fill] (6.2,9.9) circle [x radius=0.08, y radius=0.08];
	\node [] at (2.6,7.2) {\scriptsize $N_{1}$};
	\node [] at (6.8,7.1) {\scriptsize $N_{2}$};
	\node [] at (2.6,6.9) {\scriptsize $\text{edges}$};
	\node [] at (6.8,6.8) {\scriptsize $\text{edges}$};
	\end{tikzpicture}
}

\newcommand{\ExpFrobeniusAlgebraCompactAA}{
	\Biggl|\;\begin{matrix}
		\begin{tikzpicture}[scale=0.7]
		\draw [->-] (1.6,9.2) -- (2.,9.8);
		\node [] at (1.62,9.62) {\scriptsize $b$};
		\draw [->-] (1.6,10.4) -- (2.,9.8);
		\node [] at (1.62,9.98) {\scriptsize $a$};
		\draw [very thick] (2.,9.8) -- (2.8,9.8);
		\draw [->-] (2.8,9.8) -- (3.2,10.4);
		\node [] at (3.18,9.98) {\scriptsize $e$};
		\draw [->-] (3.2,9.2) -- (2.8,9.8);
		\node [] at (3.18,9.62) {\scriptsize $d$};
		\draw [fill] (2.,9.8) circle [x radius=0.08, y radius=0.08];
		\draw [fill] (2.8,9.8) circle [x radius=0.08, y radius=0.08];
		\end{tikzpicture}\:
	\end{matrix}\Biggr\rangle
}
\newcommand{\ExpFrobeniusAlgebraCompactAB}{
	\Biggl|\;\begin{matrix}
		\begin{tikzpicture}[scale=0.7]
		\draw [->-] (1.6,10.4) -- (2.4,10.2);
		\node [] at (1.95,10.09) {\scriptsize $a$};
		\draw [->-] (2.4,10.2) -- (3.2,10.4);
		\node [] at (2.85,10.09) {\scriptsize $e$};
		\draw [very thick] (2.4,9.4) -- (2.4,10.2);
		\draw [->-] (1.6,9.2) -- (2.4,9.4);
		\node [] at (1.95,9.51) {\scriptsize $b$};
		\draw [->-] (3.2,9.2) -- (2.4,9.4);
		\node [] at (2.85,9.51) {\scriptsize $d$};
		\draw [fill] (2.4,10.2) circle [x radius=0.08, y radius=0.08];
		\draw [fill] (2.4,9.4) circle [x radius=0.08, y radius=0.08];
		\end{tikzpicture}\:
	\end{matrix}\Biggr\rangle
}
\newcommand{\ExpGroundStateBoundaryComponent}{
	\Biggl|\;\begin{matrix}
		\begin{tikzpicture}[scale=1]
		\draw [lightgray, fill] (3.2,10.2) ..controls (3.558,10.2) and (3.916,10.263) .. (4.2,10.2)
		..controls (4.484,10.137) and (4.695,9.947) .. (4.8,9.6)
		..controls (4.905,9.253) and (4.905,8.747) .. (4.8,8.4)
		..controls (4.695,8.053) and (4.484,7.863) .. (4.2,7.8)
		..controls (3.916,7.737) and (3.558,7.8) .. (3.2,7.8)
		..controls (2.842,7.8) and (2.484,7.737) .. (2.2,7.8)
		..controls (1.916,7.863) and (1.705,8.053) .. (1.6,8.4)
		..controls (1.495,8.747) and (1.495,9.253) .. (1.6,9.6)
		..controls (1.705,9.947) and (1.916,10.137) .. (2.2,10.2)
		..controls (2.484,10.263) and (2.842,10.2) .. (3.2,10.2);
		\draw [white, fill] (2.4,9.) ..controls (2.4,8.771) and (2.457,8.543) .. (2.6,8.4)
		..controls (2.743,8.257) and (2.971,8.2) .. (3.2,8.2)
		..controls (3.429,8.2) and (3.657,8.257) .. (3.8,8.4)
		..controls (3.943,8.543) and (4.,8.771) .. (4.,9.)
		..controls (4.,9.229) and (3.943,9.457) .. (3.8,9.6)
		..controls (3.657,9.743) and (3.429,9.8) .. (3.2,9.8)
		..controls (2.971,9.8) and (2.743,9.743) .. (2.6,9.6)
		..controls (2.457,9.457) and (2.4,9.229) .. (2.4,9.);
		\draw [thick, dashed] (3.2,9.8) ..controls (2.967,9.767) and (2.733,9.733) .. (2.6,9.6)
		..controls (2.467,9.467) and (2.433,9.233) .. (2.4,9.);
		\draw [very thick] (2.4,9.) ..controls (2.433,8.767) and (2.467,8.533) .. (2.6,8.4)
		..controls (2.733,8.267) and (2.967,8.233) .. (3.2,8.2);
		\draw [very thick] (3.2,8.2) ..controls (3.433,8.233) and (3.667,8.267) .. (3.8,8.4)
		..controls (3.933,8.533) and (3.967,8.767) .. (4.,9.);
		\draw [very thick] (4.,9.) ..controls (3.967,9.233) and (3.933,9.467) .. (3.8,9.6)
		..controls (3.667,9.733) and (3.433,9.767) .. (3.2,9.8);
		\draw [very thick] (3.2,8.7) -- (3.2,8.2);
		\draw [very thick] (3.2,9.3) -- (3.2,9.8);
		\draw [very thick] (3.8,9.) -- (3.4,9.);
		\draw [very thick] (2.6,9.) -- (3.,9.);
		\node [fill=white,draw,rounded corners=.06cm] at (3.2,8.2) {\scriptsize $\rho_M$};
		\node [fill=white,draw,rounded corners=.06cm] at (4.,9.) {\scriptsize $\rho_M$};
		\node [fill=white,draw,rounded corners=.06cm] at (3.2,9.8) {\scriptsize $\rho_M$};
		\node [fill=white,draw,rounded corners=.06cm] at (2.4,9.) {\scriptsize $\rho_M$};
		\end{tikzpicture}\:
	\end{matrix}\Biggr\rangle
}

\newcommand{\GroundStateCylinderAA}{
	\begin{tikzpicture}[scale=1]
	\draw [dashed] (3.3,9.8) ..controls (2.992,9.783) and (2.683,9.767) .. (2.5,9.6)
	..controls (2.317,9.433) and (2.258,9.117) .. (2.2,8.8);
	\draw [->-] (2.2,8.8) ..controls (2.225,8.658) and (2.25,8.517) .. (2.3,8.4)
	..controls (2.35,8.283) and (2.425,8.192) .. (2.5,8.1);
	\draw [->-] (2.5,8.1) ..controls (2.583,8.025) and (2.667,7.95) .. (2.8,7.9)
	..controls (2.933,7.85) and (3.117,7.825) .. (3.3,7.8);
	\draw [->-] (3.3,7.8) ..controls (3.483,7.825) and (3.667,7.85) .. (3.8,7.9)
	..controls (3.933,7.95) and (4.017,8.025) .. (4.1,8.1);
	\draw [->-] (4.1,8.1) ..controls (4.175,8.192) and (4.25,8.283) .. (4.3,8.4)
	..controls (4.35,8.517) and (4.375,8.658) .. (4.4,8.8);
	\draw [->-] (4.4,8.8) ..controls (4.375,8.942) and (4.35,9.083) .. (4.3,9.2)
	..controls (4.25,9.317) and (4.175,9.408) .. (4.1,9.5);
	\draw [->-] (4.1,9.5) ..controls (4.017,9.575) and (3.933,9.65) .. (3.8,9.7)
	..controls (3.667,9.75) and (3.483,9.775) .. (3.3,9.8);
	\draw [->-] (1.6,8.8) -- (2.2,8.8);
	\node [] at (1.9,8.58) {\scriptsize $a_{1}$};
	\draw [->-] (3.3,7.2) -- (3.3,7.8);
	\node [] at (3.59,7.5) {\scriptsize $a_{2}$};
	\draw [->-] (3.3,10.4) -- (3.3,9.8);
	\node [] at (3.01,10.1) {\scriptsize $a_{4}$};
	\draw [->-] (5.,8.8) -- (4.4,8.8);
	\node [] at (4.7,9.02) {\scriptsize $a_{3}$};
	\draw [->-] (3.,8.5) -- (2.5,8.1);
	\node [] at (2.58,8.51) {\scriptsize $b_{1}$};
	\draw [->-] (3.8,8.6) -- (4.1,8.1);
	\node [] at (3.71,8.2) {\scriptsize $b_{2}$};
	\draw [->-] (3.6,9.2) -- (4.1,9.5);
	\node [] at (3.98,9.13) {\scriptsize $b_{3}$};
	\node [] at (2.,8.2) {\scriptsize $j_{1}$};
	\node [] at (2.8,7.6) {\scriptsize $j_{2}$};
	\node [] at (3.9,7.7) {\scriptsize $j_{3}$};
	\node [] at (4.6,8.3) {\scriptsize $j_{4}$};
	\node [] at (4.5,9.3) {\scriptsize $j_{5}$};
	\node [] at (3.9,9.9) {\scriptsize $j_{6}$};
	\end{tikzpicture}
}

\newcommand{\GroundStateDiskAA}{
	\begin{tikzpicture}[scale=0.7]
	\draw [lightgray, fill] (2.1,9.1) ..controls (2.1,8.871) and (2.157,8.643) .. (2.3,8.5)
	..controls (2.443,8.357) and (2.671,8.3) .. (2.9,8.3)
	..controls (3.129,8.3) and (3.357,8.357) .. (3.5,8.5)
	..controls (3.643,8.643) and (3.7,8.871) .. (3.7,9.1)
	..controls (3.7,9.329) and (3.643,9.557) .. (3.5,9.7)
	..controls (3.357,9.843) and (3.129,9.9) .. (2.9,9.9)
	..controls (2.671,9.9) and (2.443,9.843) .. (2.3,9.7)
	..controls (2.157,9.557) and (2.1,9.329) .. (2.1,9.1);
	\draw [thick, dashed] (2.9,9.9) ..controls (2.667,9.867) and (2.433,9.833) .. (2.3,9.7)
	..controls (2.167,9.567) and (2.133,9.333) .. (2.1,9.1);
	\draw [->-] (2.1,9.1) ..controls (2.133,8.867) and (2.167,8.633) .. (2.3,8.5)
	..controls (2.433,8.367) and (2.667,8.333) .. (2.9,8.3);
	\draw [->-] (2.9,8.3) ..controls (3.133,8.333) and (3.367,8.367) .. (3.5,8.5)
	..controls (3.633,8.633) and (3.667,8.867) .. (3.7,9.1);
	\draw [->-] (3.7,9.1) ..controls (3.667,9.333) and (3.633,9.567) .. (3.5,9.7)
	..controls (3.367,9.833) and (3.133,9.867) .. (2.9,9.9);
	\draw [->-] (1.6,9.1) -- (2.1,9.1);
	\node [] at (1.85,8.88) {\scriptsize $a_{N}$};
	\draw [->-] (2.9,7.8) -- (2.9,8.3);
	\node [] at (3.19,8.05) {\scriptsize $a_{1}$};
	\draw [->-] (4.2,9.1) -- (3.7,9.1);
	\node [] at (3.95,9.32) {\scriptsize $a_{2}$};
	\draw [->-] (2.9,10.4) -- (2.9,9.9);
	\node [] at (2.61,10.15) {\scriptsize $a_{3}$};
	\node [] at (2.1,8.3) {\scriptsize $l_{1}$};
	\node [] at (3.9,8.5) {\scriptsize $l_{2}$};
	\node [] at (3.7,9.9) {\scriptsize $l_{3}$};
	\end{tikzpicture}
}

\newcommand{\ExpModuleDef}{
	\Biggl|\;\begin{matrix}
		\begin{tikzpicture}[scale=0.8]
		\draw [->-] (1.6,8.) -- (1.6,8.8);
		\node [] at (1.31,8.4) {\scriptsize $j_{1}$};
		\draw [->-] (1.6,9.6) -- (1.6,10.4);
		\node [] at (1.31,10.) {\scriptsize $j_{2}$};
		\draw [->-] (2.4,8.4) -- (1.6,8.8);
		\node [] at (2.11,8.83) {\scriptsize $a_{1}$};
		\draw [->-] (2.4,9.2) -- (1.6,9.6);
		\node [] at (2.11,9.63) {\scriptsize $a_{2}$};
		\draw [very thick] (1.6,8.8) -- (1.6,9.6);
		\node [fill=white,draw,rounded corners=.06cm] at (1.6,8.8) {\scriptsize $\rho$};
		\node [fill=white,draw,rounded corners=.06cm] at (1.6,9.6) {\scriptsize $\rho$};
		\end{tikzpicture}\:
	\end{matrix}\Biggr\rangle
}
\newcommand{\ExpModuleDefAA}{
	\Biggl|\;\begin{matrix}
		\begin{tikzpicture}[scale=0.8]
		\draw [->-] (1.6,8.) -- (1.6,9.2);
		\node [] at (1.31,8.6) {\scriptsize $j_{1}$};
		\draw [->-] (1.6,9.2) -- (1.6,10.4);
		\node [] at (1.31,9.8) {\scriptsize $j_{2}$};
		\draw [->-] (2.3,8.2) -- (2.2,8.8);
		\node [] at (1.96,8.45) {\scriptsize $a_{1}$};
		\draw [->-] (2.8,8.6) -- (2.2,8.8);
		\node [] at (2.58,8.93) {\scriptsize $a_{2}$};
		\draw [very thick] (2.2,8.8) -- (1.6,9.2);
		\draw [fill] (2.2,8.8) circle [x radius=0.08, y radius=0.08];
		\node [fill=white,draw,rounded corners=.06cm] at (1.6,9.2) {\scriptsize $\rho$};
		\end{tikzpicture}\:
	\end{matrix}\Biggr\rangle
}
\newcommand{\ExpQuasiparticleBasis}{
	\Biggl|\;\begin{matrix}
		\begin{tikzpicture}[scale=0.8]
		\path [lightgray, fill] (1.6,6.4) -- (2.2,6.4) -- (2.3,6.8) -- (2.2,7.2) -- (2.3,7.6) -- (2.2,8.) -- (2.3,8.4) -- (2.2,8.8) -- (2.3,9.2) -- (2.2,9.6) -- (2.3,10.) -- (2.2,10.4) -- (1.6,10.4) -- cycle;
		\draw [] (1.9,6.4) -- (2.2,6.4);
		\draw [] (1.9,7.2) -- (2.2,7.2);
		\draw [] (1.9,8.) -- (2.2,8.);
		\draw [] (1.9,8.8) -- (2.2,8.8);
		\draw [] (1.9,9.6) -- (2.2,9.6);
		\draw [] (1.9,10.4) -- (2.2,10.4);
		\draw [->-] (2.2,6.4) -- (2.3,6.8);
		\draw [->-] (2.2,7.2) -- (2.3,7.6);
		\draw [->-] (2.2,8.) -- (2.3,8.4);
		\draw [->-] (2.2,8.8) -- (2.3,9.2);
		\draw [->-] (2.2,9.6) -- (2.3,10.);
		\draw [->-] (2.3,6.8) -- (2.2,7.2);
		\draw [->-] (2.3,7.6) -- (2.2,8.);
		\draw [->-] (2.3,8.4) -- (2.2,8.8);
		\draw [->-] (2.3,9.2) -- (2.2,9.6);
		\draw [->-] (2.3,10.) -- (2.2,10.4);
		\draw [->-] (2.3,6.8) -- (2.7,6.8);
		\node [] at (2.5,6.58) {\scriptsize $a_{1}$};
		\draw [->-] (2.3,7.6) -- (2.7,7.6);
		\node [] at (2.5,7.38) {\scriptsize $a_{2}$};
		\draw [->-] (2.3,8.4) -- (2.7,8.4);
		\node [] at (2.5,8.18) {\scriptsize $a_{3}$};
		\draw [->-] (2.3,9.2) -- (2.7,9.2);
		\node [] at (2.5,8.98) {\scriptsize $a_{4}$};
		\draw [->-] (2.3,10.) -- (2.7,10.);
		\node [] at (2.5,9.78) {\scriptsize $a_{5}$};
		\end{tikzpicture}\:
	\end{matrix}\Biggr\rangle
}
\newcommand{\ExpQuasiparticleBasisAA}{
	\Biggl|\;\begin{matrix}
		\begin{tikzpicture}[scale=1]
		\path [lightgray, fill] (1.6,8.8) -- (2.2,8.8) -- (2.3,9.2) -- (2.2,9.6) -- (2.3,10.) -- (2.2,10.4) -- (1.6,10.4) -- cycle;
		\draw [] (1.9,8.8) -- (2.2,8.8);
		\draw [] (1.9,9.6) -- (2.2,9.6);
		\draw [] (1.9,10.4) -- (2.2,10.4);
		\draw [->-] (2.2,8.8) -- (2.3,9.2);
		\draw [->-] (2.2,9.6) -- (2.3,10.);
		\draw [->-] (2.3,9.2) -- (2.2,9.6);
		\draw [->-] (2.3,10.) -- (2.2,10.4);
		\draw [->-] (2.3,9.2) -- (2.7,9.2);
		\node [] at (2.5,8.98) {\scriptsize $j$};
		\draw [->-] (2.3,10.) -- (2.7,10.);
		\node [] at (2.5,9.78) {\scriptsize $a$};
		\end{tikzpicture}\:
	\end{matrix}\Biggr\rangle
}
\newcommand{\ExpQuasiparticleCreation}{
	\Biggl|\;\begin{matrix}
		\begin{tikzpicture}[scale=0.8]
		\path [lightgray, fill] (1.6,6.4) -- (2.2,6.4) -- (2.3,6.8) -- (2.2,7.2) -- (2.3,7.6) -- (2.2,8.) -- (2.3,8.4) -- (2.2,8.8) -- (2.3,9.2) -- (2.2,9.6) -- (2.3,10.) -- (2.2,10.4) -- (1.6,10.4) -- cycle;
		\draw [->-] (2.3,6.8) ..controls (2.533,6.867) and (2.767,6.933) .. (2.9,7.)
		..controls (3.033,7.067) and (3.067,7.133) .. (3.1,7.2);
		\draw [->-] (2.3,7.6) ..controls (2.533,7.533) and (2.767,7.467) .. (2.9,7.4)
		..controls (3.033,7.333) and (3.067,7.267) .. (3.1,7.2);
		\draw [->-] (2.3,9.2) ..controls (2.533,9.267) and (2.767,9.333) .. (2.9,9.4)
		..controls (3.033,9.467) and (3.067,9.533) .. (3.1,9.6);
		\draw [->-] (2.3,10.) ..controls (2.533,9.933) and (2.767,9.867) .. (2.9,9.8)
		..controls (3.033,9.733) and (3.067,9.667) .. (3.1,9.6);
		\draw [very thick] (3.,8.4) ..controls (2.992,8.542) and (2.983,8.683) .. (3.1,8.8)
		..controls (3.217,8.917) and (3.458,9.008) .. (3.7,9.1);
		\draw [very thick] (3.7,7.7) ..controls (3.458,7.792) and (3.217,7.883) .. (3.1,8.)
		..controls (2.983,8.117) and (2.992,8.258) .. (3.,8.4);
		\draw [] (1.9,6.4) -- (2.2,6.4);
		\draw [] (1.9,7.2) -- (2.2,7.2);
		\draw [] (1.9,8.) -- (2.2,8.);
		\draw [] (1.9,8.8) -- (2.2,8.8);
		\draw [] (1.9,9.6) -- (2.2,9.6);
		\draw [] (1.9,10.4) -- (2.2,10.4);
		\draw [->-] (2.2,6.4) -- (2.3,6.8);
		\draw [->-] (2.2,7.2) -- (2.3,7.6);
		\draw [->-] (2.2,8.) -- (2.3,8.4);
		\draw [->-] (2.2,8.8) -- (2.3,9.2);
		\draw [->-] (2.2,9.6) -- (2.3,10.);
		\draw [->-] (2.3,6.8) -- (2.2,7.2);
		\draw [->-] (2.3,7.6) -- (2.2,8.);
		\draw [->-] (2.3,8.4) -- (2.2,8.8);
		\draw [->-] (2.3,8.4) -- (2.8,8.4);
		\draw [->-] (2.3,9.2) -- (2.2,9.6);
		\draw [->-] (2.3,10.) -- (2.2,10.4);
		\draw [very thick] (3.3,8.4) -- (3.7,8.4);
		\draw [very thick] (3.1,7.2) -- (3.7,7.2);
		\draw [very thick] (3.1,9.6) -- (3.7,9.6);
		\draw [fill] (3.1,7.2) circle [x radius=0.08, y radius=0.08];
		\draw [fill] (3.1,9.6) circle [x radius=0.08, y radius=0.08];
		\node [fill=white,draw,rounded corners=.06cm] at (3.,8.4) {\scriptsize $P_M$};
		\node [] at (2.7,6.7) {\scriptsize $a_{1}$};
		\node [] at (2.6,7.3) {\scriptsize $a_{2}$};
		\node [] at (2.6,8.) {\scriptsize $a_{3}$};
		\node [] at (2.7,9.1) {\scriptsize $a_{4}$};
		\node [] at (2.7,10.1) {\scriptsize $a_{5}$};
		\end{tikzpicture}\:
	\end{matrix}\Biggr\rangle
}

\newcommand{\ExpQuasiparticleHoppingAA}{
	\Biggl|\;\begin{matrix}
		\begin{tikzpicture}[scale=1]
		\path [lightgray, fill] (1.6,8.8) -- (2.2,8.8) -- (2.3,9.2) -- (2.2,9.6) -- (2.3,10.) -- (2.2,10.4) -- (1.6,10.4) -- cycle;
		\draw [->-] (2.3,9.2) ..controls (2.492,9.267) and (2.683,9.333) .. (2.8,9.4)
		..controls (2.917,9.467) and (2.958,9.533) .. (3.,9.6);
		\draw [->-] (2.3,10.) ..controls (2.492,9.933) and (2.683,9.867) .. (2.8,9.8)
		..controls (2.917,9.733) and (2.958,9.667) .. (3.,9.6);
		\draw [very thick] (3.,9.6) ..controls (3.05,9.75) and (3.1,9.9) .. (3.2,10.)
		..controls (3.3,10.1) and (3.45,10.15) .. (3.6,10.2);
		\draw [very thick] (3.,9.6) ..controls (3.05,9.492) and (3.1,9.383) .. (3.2,9.3)
		..controls (3.3,9.217) and (3.45,9.158) .. (3.6,9.1);
		\draw [] (1.9,8.8) -- (2.2,8.8);
		\draw [] (1.9,9.6) -- (2.2,9.6);
		\draw [] (1.9,10.4) -- (2.2,10.4);
		\draw [->-] (2.2,8.8) -- (2.3,9.2);
		\draw [->-] (2.2,9.6) -- (2.3,10.);
		\draw [->-] (2.3,9.2) -- (2.2,9.6);
		\draw [->-] (2.3,10.) -- (2.2,10.4);
		\node [fill=white,draw,rounded corners=.06cm] at (3.,9.6) {\scriptsize $P_M$};
		\node [] at (2.6,9.) {\scriptsize $j$};
		\node [] at (2.7,10.1) {\scriptsize $a$};
		\end{tikzpicture}\:
	\end{matrix}\Biggr\rangle
}
\newcommand{\ExpStrongFrobeniusAB}{
	\Biggl|\;\begin{matrix}
		\begin{tikzpicture}[scale=0.7]
		\draw [->-] (1.6,8.4) -- (1.6,10.4);
		\node [] at (1.82,9.4) {\scriptsize $c$};
		\end{tikzpicture}\:
	\end{matrix}\Biggr\rangle
}
\newcommand{\ExpStrongFrobeniusCompactAA}{
	\Biggl|\;\begin{matrix}
		\begin{tikzpicture}[scale=0.7]
		\draw [very thick] (2.2,9.) ..controls (2.44,9.044) and (2.68,9.089) .. (2.8,9.2)
		..controls (2.92,9.311) and (2.92,9.489) .. (2.8,9.6)
		..controls (2.68,9.711) and (2.44,9.756) .. (2.2,9.8);
		\draw [very thick] (2.2,9.) ..controls (1.96,9.044) and (1.72,9.089) .. (1.6,9.2)
		..controls (1.48,9.311) and (1.48,9.489) .. (1.6,9.6)
		..controls (1.72,9.711) and (1.96,9.756) .. (2.2,9.8);
		\draw [->-] (2.2,9.8) -- (2.2,10.4);
		\node [] at (2.42,10.1) {\scriptsize $c$};
		\draw [->-] (2.2,8.4) -- (2.2,9.);
		\node [] at (2.42,8.7) {\scriptsize $c$};
		\draw [fill] (2.2,9.) circle [x radius=0.08, y radius=0.08];
		\draw [fill] (2.2,9.8) circle [x radius=0.08, y radius=0.08];
		\end{tikzpicture}\:
	\end{matrix}\Biggr\rangle
}
\newcommand{\TailLabel}[3]{
	\begin{tikzpicture}[scale=0.8]
	\path [lightgray, fill] (1.6,8.8) -- (2.2,8.8) -- (2.2,9.6) -- (2.2,10.4) -- (1.6,10.4) -- cycle;
	\draw [->-] (2.2,8.8) -- (2.2,9.6);
	\draw [->-] (2.2,9.6) -- (2.2,10.4);
	\draw [->-] (2.2,9.6) -- (2.8,9.6);
	\node [] at (2.6,9.4) {\scriptsize $#1$};
	\node [] at (1.8,9.2) {\scriptsize $#2$};
	\node [] at (1.8,10.) {\scriptsize $#3$};
	\end{tikzpicture}
}

\newcommand{\ExpTBdryMoveOneTwo}{
	\Biggl|\;\begin{matrix}
		\begin{tikzpicture}[scale=1]
		\path [lightgray, fill] (1.6,8.8) -- (2.2,8.8) -- (2.2,9.6) -- (2.2,10.4) -- (1.6,10.4) -- cycle;
		\draw [->-] (2.2,9.6) -- (2.8,9.6);
		\node [] at (2.5,9.38) {\scriptsize $a_{1}$};
		\draw [->-] (2.2,8.8) -- (2.2,9.6);
		\node [] at (1.98,9.2) {\scriptsize $i$};
		\draw [->-] (2.2,9.6) -- (2.2,10.4);
		\node [] at (1.98,10.) {\scriptsize $j$};
		\end{tikzpicture}\:
	\end{matrix}\Biggr\rangle
}
\newcommand{\ExpTBdryMoveOneTwoAC}{
	\sum_{a_{2},a_{3}}\frac{\mathrm{u}_{a_{1}}\mathrm{u}_{a_{2}}\mathrm{u}_{a_{3}}}{\sqrt{\mathrm{d}_A}}\sum_{k}\mathrm{v}_{k}f_{a_{2}^* a_{3}^* a_{1}}G^{j^* i a_{1}^*}_{a_{2}^* a_{3}^* k}
	\Biggl|\;\begin{matrix}
		\begin{tikzpicture}[scale=1]
		\path [lightgray, fill] (1.6,8.8) -- (2.2,8.8) -- (2.2,9.6) -- (2.2,10.4) -- (1.6,10.4) -- cycle;
		\draw [->-] (2.2,8.8) -- (2.2,9.3);
		\node [] at (1.98,9.05) {\scriptsize $i$};
		\draw [->-] (2.2,9.3) -- (2.2,9.9);
		\node [] at (1.98,9.6) {\scriptsize $k$};
		\draw [->-] (2.2,9.3) -- (3.,9.3);
		\node [] at (2.6,9.08) {\scriptsize $a_{2}$};
		\draw [->-] (2.2,9.9) -- (2.2,10.4);
		\node [] at (1.98,10.15) {\scriptsize $j$};
		\draw [->-] (2.2,9.9) -- (3.,9.9);
		\node [] at (2.6,10.12) {\scriptsize $a_{3}$};
		\end{tikzpicture}\:
	\end{matrix}\Biggr\rangle
}
\newcommand{\ExpTBdryMoveTwoOne}{
	\Biggl|\;\begin{matrix}
		\begin{tikzpicture}[scale=1]
		\path [lightgray, fill] (1.6,8.8) -- (2.2,8.8) -- (2.2,9.6) -- (2.2,10.4) -- (1.6,10.4) -- cycle;
		\draw [->-] (2.2,8.8) -- (2.2,9.3);
		\node [] at (1.98,9.05) {\scriptsize $i$};
		\draw [->-] (2.2,9.3) -- (2.2,9.9);
		\node [] at (1.98,9.6) {\scriptsize $k$};
		\draw [->-] (2.2,9.3) -- (3.,9.3);
		\node [] at (2.6,9.08) {\scriptsize $a_{2}$};
		\draw [->-] (2.2,9.9) -- (2.2,10.4);
		\node [] at (1.98,10.15) {\scriptsize $j$};
		\draw [->-] (2.2,9.9) -- (3.,9.9);
		\node [] at (2.6,10.12) {\scriptsize $a_{3}$};
		\end{tikzpicture}\:
	\end{matrix}\Biggr\rangle
}
\newcommand{\ExpTBdryMoveTwoOneAC}{
	\sum_{a_{1}}\frac{\mathrm{u}_{a_{1}}\mathrm{u}_{a_{2}}\mathrm{u}_{a_{3}}}{\sqrt{\mathrm{d}_A}}\mathrm{v}_{k}f_{a_{2} a_{3} a_{1}^*}G^{j a_{1} i^*}_{a_{2}^* k^* a_{3}}
	\Biggl|\;\begin{matrix}
		\begin{tikzpicture}[scale=1]
		\path [lightgray, fill] (1.6,8.8) -- (2.2,8.8) -- (2.2,9.6) -- (2.2,10.4) -- (1.6,10.4) -- cycle;
		\draw [->-] (2.2,8.8) -- (2.2,9.6);
		\node [] at (1.98,9.2) {\scriptsize $i$};
		\draw [->-] (2.2,9.6) -- (2.2,10.4);
		\node [] at (1.98,10.) {\scriptsize $j$};
		\draw [->-] (2.2,9.6) -- (3.,9.6);
		\node [] at (2.6,9.38) {\scriptsize $a_{1}$};
		\end{tikzpicture}\:
	\end{matrix}\Biggr\rangle
}
\newcommand{\ExpTMoveOneThree}{
	\Biggl|\;\begin{matrix}
		\begin{tikzpicture}[scale=0.7]
		\draw [->-] (1.6,10.4) -- (2.2,9.8);
		\node [] at (1.71,9.91) {\scriptsize $j_{1}$};
		\draw [->-] (2.2,9.) -- (2.2,9.8);
		\node [] at (2.49,9.4) {\scriptsize $j_{2}$};
		\draw [->-] (2.8,10.4) -- (2.2,9.8);
		\node [] at (2.69,9.91) {\scriptsize $j_{3}$};
		\end{tikzpicture}\:
	\end{matrix}\Biggr\rangle
}
\newcommand{\ExpTMoveOneThreeAA}{
	\sum_{j_{4},j_{5},j_{6},}\frac{\mathrm{v}_{j_{5}}\mathrm{v}_{j_{6}}\mathrm{v}_{j_{4}}}{\sqrt{D}}G^{j_{1} j_{2} j_{3}}_{j_{5} j_{6} j_{4}}
	\Biggl|\;\begin{matrix}
		\begin{tikzpicture}[scale=0.7]
		\draw [->-] (1.6,10.4) -- (1.9,10.1);
		\node [] at (1.56,10.06) {\scriptsize $j_{1}$};
		\draw [->-] (2.3,9.) -- (2.3,9.4);
		\node [] at (2.59,9.2) {\scriptsize $j_{2}$};
		\draw [->-] (2.3,9.4) -- (1.9,10.1);
		\node [] at (1.85,9.61) {\scriptsize $j_{4}$};
		\draw [->-] (2.7,10.1) -- (1.9,10.1);
		\node [] at (2.3,10.32) {\scriptsize $j_{6}$};
		\draw [->-] (3.,10.4) -- (2.7,10.1);
		\node [] at (3.04,10.06) {\scriptsize $j_{3}$};
		\draw [->-] (2.7,10.1) -- (2.3,9.4);
		\node [] at (2.75,9.61) {\scriptsize $j_{5}$};
		\end{tikzpicture}\:
	\end{matrix}\Biggr\rangle
}
\newcommand{\ExpTMoveThreeOne}{
	\Biggl|\;\begin{matrix}
		\begin{tikzpicture}[scale=0.7]
		\draw [->-] (1.6,10.4) -- (1.9,10.1);
		\node [] at (1.56,10.06) {\scriptsize $j_{1}$};
		\draw [->-] (2.3,9.) -- (2.3,9.4);
		\node [] at (2.59,9.2) {\scriptsize $j_{2}$};
		\draw [->-] (2.3,9.4) -- (1.9,10.1);
		\node [] at (1.85,9.61) {\scriptsize $j_{4}$};
		\draw [->-] (2.7,10.1) -- (1.9,10.1);
		\node [] at (2.3,10.32) {\scriptsize $j_{6}$};
		\draw [->-] (3.,10.4) -- (2.7,10.1);
		\node [] at (3.04,10.06) {\scriptsize $j_{3}$};
		\draw [->-] (2.7,10.1) -- (2.3,9.4);
		\node [] at (2.75,9.61) {\scriptsize $j_{5}$};
		\end{tikzpicture}\:
	\end{matrix}\Biggr\rangle
}
\newcommand{\ExpTMoveThreeOneAA}{
	\frac{\mathrm{v}_{j_{5}}\mathrm{v}_{j_{4}}\mathrm{v}    _{j_{6}}}{\sqrt{D}}G^{j_{1}^* j_{3}^* j_{2}^*}_{j_{5} j_{4}^* j_{6}^*}
	\Biggl|\;\begin{matrix}
		\begin{tikzpicture}[scale=0.7]
		\draw [->-] (1.6,10.4) -- (2.3,9.8);
		\node [] at (1.78,9.9) {\scriptsize $j_{1}$};
		\draw [->-] (2.3,9.) -- (2.3,9.8);
		\node [] at (2.59,9.4) {\scriptsize $j_{2}$};
		\draw [->-] (3.,10.4) -- (2.3,9.8);
		\node [] at (2.82,9.9) {\scriptsize $j_{3}$};
		\end{tikzpicture}\:
	\end{matrix}\Biggr\rangle
}
\newcommand{\ExpTMoveTwoTwo}{
	\Biggl|\;\begin{matrix}
		\begin{tikzpicture}[scale=0.7]
		\draw [->-] (1.6,10.4) -- (2.,9.6);
		\node [] at (1.54,9.87) {\scriptsize $j_{1}$};
		\draw [->-] (2.8,9.6) -- (2.,9.6);
		\node [] at (2.4,9.82) {\scriptsize $j_{5}$};
		\draw [->-] (3.2,8.8) -- (2.8,9.6);
		\node [] at (3.26,9.33) {\scriptsize $j_{3}$};
		\draw [->-] (3.2,10.4) -- (2.8,9.6);
		\node [] at (2.74,10.13) {\scriptsize $j_{4}$};
		\draw [->-] (1.6,8.8) -- (2.,9.6);
		\node [] at (1.54,9.33) {\scriptsize $j_{2}$};
		\end{tikzpicture}\:
	\end{matrix}\Biggr\rangle
}
\newcommand{\ExpTMoveTwoTwoAA}{
	\sum_{j'_{5}}G^{j_{1} j_{2} j_{5}}_{j_{3} j_{4} j'_{5}}\mathrm{v}_{j_{5}}\mathrm{v}_{j'_{5}}
	\Biggl|\;\begin{matrix}
		\begin{tikzpicture}[scale=0.7]
		\draw [->-] (1.6,10.4) -- (2.4,10.);
		\node [] at (1.89,9.97) {\scriptsize $j_{1}$};
		\draw [->-] (2.4,9.2) -- (2.4,10.);
		\node [] at (2.77,9.6) {\scriptsize $j'_{5}$};
		\draw [->-] (3.2,10.4) -- (2.4,10.);
		\node [] at (2.91,9.97) {\scriptsize $j_{4}$};
		\draw [->-] (1.6,8.8) -- (2.4,9.2);
		\node [] at (1.89,9.23) {\scriptsize $j_{2}$};
		\draw [->-] (3.2,8.8) -- (2.4,9.2);
		\node [] at (2.91,9.23) {\scriptsize $j_{3}$};
		\end{tikzpicture}\:
	\end{matrix}\Biggr\rangle
}

\begin{document}
	
	\title{Boundary Hamiltonian theory for gapped topological orders} 
	
	\date{\today}

	\author{Yuting Hu}
	\affiliation{Department of Physics and Center for Field Theory and Particle Physics, Fudan University, Shanghai 200433, China}
	\email{yuting.phys@gmail.com}
	\author{Yidun Wan}
	\affiliation{Department of Physics and Center for Field Theory and Particle Physics, Fudan University, Shanghai 200433, China}
	\affiliation{Collaborative Innovation Center of Advanced Microstructures, Nanjing 210093, China}
	\email{ydwan@fudan.edu.cn}
	\author{Yong-Shi Wu}
	\affiliation{State Key Laboratory of Surface Physics, Fudan University, Shanghai 200433, China}
	\affiliation{Department of Physics and Center for Field Theory and Particle Physics, Fudan University, Shanghai 200433, China}
	\affiliation{Collaborative Innovation Center of Advanced Microstructures, Nanjing 210093, China}
	\affiliation{Department of Physics and Astronomy, University of Utah, Salt Lake City, Utah, 84112, U.S.A.}
	\email{yswu@fudan.eud.cn}

	\begin{abstract}
		In this letter, we report our systematic construction of the lattice Hamiltonian  model of topological orders on open surfaces, with explicit boundary terms. We do this mainly for the Levin-Wen string-net model. The full Hamiltonian in our approach yields a topologically protected, gapped energy spectrum, with the corresponding wave functions robust under topology-preserving transformations of the lattice of the system. We explicitly present the wavefunctions of the ground states and boundary elementary excitations. We construct the creation and hopping operators of boundary quasi-particles. We find that given a bulk topological order, the gapped boundary conditions are classified by Frobenius algebras in its input data. Emergent topological properties of the ground states and boundary excitations are characterized by (bi-) modules over Frobenius algebras.

	\end{abstract} 
	
	
	\maketitle

	{\bf Introduction.} Matter phases with intrinsic topological orders (ITO) not only extend our understanding of phases of matter far beyond the Landau-Ginzburg paradigm\cite{Wen1989,Wen1991} but also may support robust quantum memories\cite{Dennis2002} and topological quantum computation\cite{Kitaev2003a,Freedman2003,Stern2006,Nayak2008}. Most studies of the dynamical theories of 2d ITO have been unfortunately limited to closed surfaces (except Ref.  [\onlinecite{Beigi2011,IMZ2016,IMZ2017}] for the Kitaev model). On the one hand, closed surface materials are hardly realizable in experiments. This situation hinders the realistic applicability of ITO. On the other hand, a bulk Hamiltonian theory is incomplete for a system on open surfaces, if its boundary conditions are not specified. In this letter, we report our explicit and systematic construction of a wide class of boundary Hamiltonians for string-net models, to specify the boundary conditions and to be added to the bulk Levin-Wen bulk Hamiltonian\cite{Levin2004}. We present explicit ground-state wavefunctions and construct creation and hopping operators of boundary excitations in our new approach. (We have also systematically constructed the boundary Hamiltonian of the twisted quantum double model\cite{Hu2012a} of ITO, which is to be reported elsewhere.)  
	
	We shall make heavy use of the graphic techniques, of which details can be found in original references\cite{Levin2004,Hu2012}. With the graphic rules, a graph equality can be unambiguously turned into usual algebraic equations involving tensors. Our notations follow the conventions in [\onlinecite{Hu2012}]. 
	
	{\bf Preliminaries.} Non-chiral bulk intrinsic topological phases in (2+1)-D can be studied by effective discrete topological quantum field theories, such as the LW model, an exactly solvable Hamiltonian model defined on 2d spatial trivalent graphs. The LW model is essentially a discrete gauge theory, defined using a unitary fusion category (UFC) $\mathcal{C}$, e.g. the collection of all representations of a finite or quantum group, as input data (see, e.g., Ref.\onlinecite{Wu2016}). For simplicity, we assume $\mathcal{C}$ is multiplicity free. 
	The Hamiltonian is defined in terms of the $6j$-symbols over $\mathcal{C}$. The topological properties of the LW model, such as the ground state degeneracy (GSD) and the topological quantum numbers of the quasi-particle excitations are ensured by their invariance under the change of $\Gamma$ by the Pachner moves. There are three elementary Pachner moves, associated with which are respectively three unitary linear maps\cite{Hu2012} in the ground-state subspace :
	\begin{align}
	\label{T1T2T3}
	&T_{2\rightarrow 2}\ExpTMoveTwoTwo
	=\ExpTMoveTwoTwoAA\nonumber\\
	&T_{1\rightarrow 3}\ExpTMoveOneThree
	=\ExpTMoveOneThreeAA\nonumber\\
	&T_{3\rightarrow 1}\ExpTMoveThreeOne
	=\ExpTMoveThreeOneAA,
	\end{align}
	where $j_i$'s are inequivalent simple objects (usually called string-types) of $\mathcal{C}$, $G$'s are the symmetric $6j$-symbols\cite{Hu2012} over $\mathcal{C}$ with normalization $v_j=1/G^{j^*j0}_{00j}$, and $D=\sum_j v_j^4$. 	These operators yield a unique transformation between the Hilbert spaces before and after the Pachner moves. We shall denote the unique transformation by $\mathcal{T}$, which is independent of the path of Pachner moves involved. 
	
	{\bf Frobenius algebra and boundary Hamiltonian.} We propose to use the Frobenius algebras in a UFC to specify boundary conditions and construct boundary terms to be added to the LW Hamiltonian. Let $G$ be a symmetric $6j$-symbol over the string-type set $L$, the set of all (inequivalent) simple objects of $\mathcal{C}$. A \textit{Frobenius algebra} $A$ (in $\mathcal{C}$) is a subset $L_A$ of $L$ equipped with a multiplication $f_{abc^*}$, satisfying
	\begin{align}
	\label{eq:fcond}
	&\text{(associativity)}  \sum_{c}f_{abc^*}f_{cde^*}G^{abc^*}_{de^*g}\rmv_c\rmv_g,
	=f_{age^*}f_{bdg^*},\nonumber\\
	&\text{(non-degeneracy)}\quad   f_{bb^*0}\neq 0, \quad\forall b\in L_A,
	\end{align}
	where all indices take values in $L_A$. Due to the symmetry conditions of the symmetric $6j$-symbols\cite{Hu2012}, the multiplication have the following defining properties.
	\begin{align}\label{eq:fsymmetry}
	&\text{(unit)}\quad f_{bb^*0}=f_{b0b^*}=f_{0bb^*}=1,\nonumber\\
	&\text{(cyclic)}\quad f_{abc}=f_{cab},\nonumber\\
	&\text{(strong)}\quad \sum_{ab}f_{abc}f_{c^*b^*a^*}\rmv_a\rmv_b=\rmd_A\rmv_c,
	\end{align}
	where $\rmd_A=\sum_{a\in L_A}\rmd_a$ is the quantum dimension of $A$.
	
	We can express the associative and strong conditions in a compact way graphically:
	\begin{equation}\label{eq:FrobeniusAlgebraCompact}
	\Tact\ExpFrobeniusAlgebraCompactAA\;=\;
	\ExpFrobeniusAlgebraCompactAB
	\end{equation}
	\begin{equation}\label{eq:StrongConditionCompact}
	\Tact{\frac{\sqrt{D}}{\rmd_A}\ExpStrongFrobeniusCompactAA}
	\;=\;
	\ExpStrongFrobeniusAB,
	\end{equation}
	Here each vertex carries a multiplication, and an unlabeled thick line implies a summation over its label in $L_A$. We take this convention in graphical equations throughout the letter.
	
	For simplicity, let us consider a system on an open surface, outside of which is the vacuum. To each boundary (component) A, we associate a Frobenius algebra $L_A$, and attach an open edge (or a tail), labeled by a string-type $a\in L_A$, from the vacuum side to each boundary edge. See Fig. 1(a), in which the grey region represents the bulk of the graph, including the edges along the rim of the grey region. Throughout the paper, we shall adopt this convention.
	
	Similar to transformations \eqref{T1T2T3}, we can use the Frobenius algebra $A$ to define unitary transformations associated with 1D Pachner moves on the boundaries of a graph: (with $u_a=\sqrt{v_a}$)         
	\begin{align}\label{eq:TMoveBoundary}
	&T_{1\rightarrow 2} \ExpTBdryMoveOneTwo
	\nonumber\\
	=&\ExpTBdryMoveOneTwoAC\nonumber\\
	&T_{2\rightarrow 1} \ExpTBdryMoveTwoOne\nonumber\\
	=&\ExpTBdryMoveTwoOneAC.
	\end{align} 
	where $\rmu_a=\sqrt{\rmv_a}$ (sign of square root may be arbitrarily chosen but if fixed once then for all).
	
	\begin{figure}[h!]
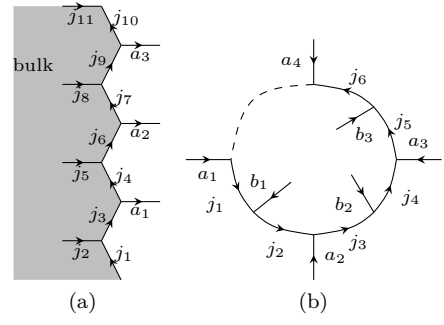

		\centering
		\subfigure[]{\FigBoundary \label{fig:Boundary}}
		\subfigure[]{\GroundStateCylinderAA	\label{fig:GroundStateCylinder}}
		\caption{(a) Boundary is a wall carrying open edges $a$'s. Bulk edges are the $j$'s. The bulk edges are not open edges, and they appear so because we neglected the rest of the bulk. (b) When no quasiparticle exists in bulk, a cylindrical system can be effecitvely described by a circular wall with all bulk plaquettes eliminated by Pachner moves.}
	\end{figure}
	
	With the boundaries, the topological feature of the entire model is described as follows. The ground-state Hilbert space is invariant under any transformation composed of $T_{2\rightarrow 2},T_{1\rightarrow 3},T_{3\rightarrow 1}$ in the bulk and $\Tmv{1}{2},\Tmv{2}{1}$ on the boundary. Associativity ensures that a product of such transformations is unique. We shall show the uniqueness for boundary Pachner moves $\Tmv{1}{2},\Tmv{2}{1}$, as that for bulk Pachner moves shown in Ref.\cite{Hu2012}. 
	
\begin{figure}
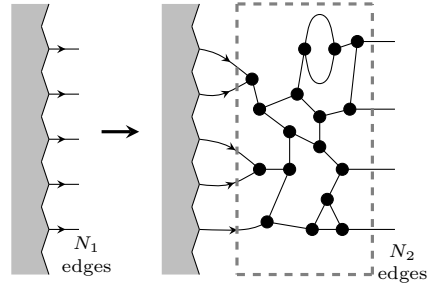

		\centering
		\FigBoundaryTopoInvarianceAA
		\caption{Sketch of proving the uniqueness of the transformation associated with boundary Pachner moves.}
		\label{fig:BoundaryTransform}
	\end{figure}

	Without loss of generality, consider the transformation from $N_1$ open edges to $N_2$ open edges (See Fig. \ref{fig:BoundaryTransform}). The composition of $\Tmv{1}{2}$ and $\Tmv{2}{1}$ amounts a graph structure with $N_1$ input edges and $N_2$ output edges, where each trivalent vertex is attached with a multiplication. From the associativity condition, the transformation presented by the graph in the dashed box is unique.

	In ground states, the boundary degrees of freedom are restricted to $L_A$. The boundary Hamiltonian reads
	\begin{equation}\label{eq:bdryHamiltonian}
	H_{\text{bdry}}=-\sum_n\Qb_n-\sum_p \Bb_{p},
	\quad
	\Bb_{p}=\dfrac{1}{\rmd_A}\sum_{t\in L_A}\rmv_t \Bb^t_{p}.
	\end{equation}
	Here, $\Qb_n$ is a boundary edge operator acting on open edge $n$, which projects the boundary d.o.f. to $L_A\subseteq L$: 
	\begin{equation}\label{eq:bdryQn}
	\Qb_n\bket{\TailLabel{a_n}{j_1}{j_2}}=\delta_{a_n\in L_A}\bket{\TailLabel{a_n}{j_1}{j_2}}.
	\end{equation}
And $\Bb_p^t$ acts on a boundary open plaquette between a pair of nearest neighboring open edges:

\begin{align}
&\Bb^t_{p}\ExpBoundaryBbpAA
=
\sum_{a'_{1}a'_{2}j'_{2}j'_{3}}
f_{t^* {a'_{2}}^* a_{2}} f_{a_{1} t {a'_{1}}^*}\mathrm{u}_{a_{1}}
\mathrm{u}_{a_{2}}\mathrm{u}_{a'_{1}}
\mathrm{u}_{a'_{2}}\times
\nonumber\\
&\,
\mathrm{v}_{j_{2}}\mathrm{v}_{j_{3}}\mathrm{v}_{j'_{2}}\mathrm{v}_{j'_{3}}
G^{j_{4}^* j_{3} a_{2}^*}_{t^* {a'_{2}}^* j'_{3}}
G^{j_{5} j_{2} j_{3}^*}_{t^* {j'_{3}}^* j'_{2}}
G^{t^* {j'_{2}}^* j_{2}}_{j_{1} a_{1}^* a'_{1}}
\bket{\BoundaryBbpAF}.
\label{eq:BptExact}
\end{align}

	Following the same convention as in eq. \eqref{eq:StrongConditionCompact},
	We can write $\Bb_p$ in a more compact fashion as
	\begin{equation}\label{eq:BbpIsTmoves}
	\Bb_p \ExpBoundaryBbpAA
	= \mathcal{T} \ExpBoundaryBbpAG
	\end{equation}
	The marked ($\times$) plaquette will be annihilated by the boundary Pachner moves to generate the coefficients in Eq. \eqref{eq:BptExact}.
	
	Boundary terms $\Qb_v$ and $\Bb_p$ are shown to be projections commuting with bulk terms $Q_{v'},B_{p'}$ and other boundary terms $\Qb_{v''},\Bb_{p''}$. Correspondence properties between bulk and boundary operators are summarized in Table. \ref{tab:BulkBoundaryHamiltonian}.
	
	\begin{table}[!ht]
		\centering
		\begin{tabular}{lll}
			\hline
			Bulk &  & Boundary \\ 
			\hline
			$B_{p=\bigtriangleup}=\Tmv{1}{3}\cdot\Tmv{3}{1}$ & & $\Bb_p=\Tmv{1}{2}\cdot\Tmv{2}{1}$ \\
			$B_p=\frac{1}{D}\sum_s \rmd_s B_p^s$ & & $\Bb_p=\frac{1}{\rmd_A}\sum_a\rmv_a \Bb_p^a$ \\
			$B_p^rB_p^s=\sum_t \delta_{rst^*}B_p^t$ &  & $\Bb_p^a \Bb_p^b=\sum_c\delta_{abc^*}\Bb_p^c$\\
			\hline
		\end{tabular}
		\caption{Correspondence between Bulk and boundary operators}
		\label{tab:BulkBoundaryHamiltonian}
	\end{table}
	
	{\bf Ground states.} A (right) module over Frobenius algebra $A$ is a tensor $\rho^a_{j_1j_2}$, with $a\in L_A$ and $j_1,j_2\in L$, satisfying
	\begin{equation}\label{eq:ModuleDef}
	\Tmv{2}{2}\ExpModuleDef=
	\ExpModuleDefAA
	\end{equation}
	Here a box $\rho$ at a vertex means that the tensor $\rho$ is associated with the vertex (e.g., $\rho^c_{j_1j_2}$ on RHS, with a summation over $c$). We denote all modules over $A$ by $Mod_{A}$.
	
	\begin{figure}[h!]
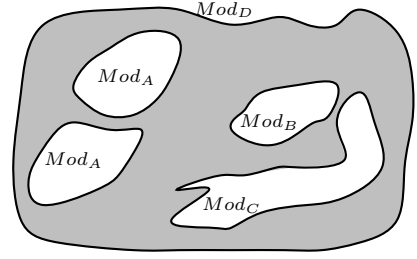

		\centering
		\FigBoundaryModuleAA
		\caption{A surface with multiple disconnected boundaries. An $A$-boundary has a $Mod_A$ ground state basis. 
		}
		\label{fig:multipleComponents}
	\end{figure}
	
	The local ground states on a boundary component $A$ is characterized by $Mod_A$. See Fig. \ref{fig:multipleComponents}. The basis (for $M\in Mod_A$) is given by
	\begin{equation}
	\label{eq:GroundStateEachComponent}
	\ket{\Phi_M}=\ExpGroundStateBoundaryComponent.
	\end{equation}
	Nevertheless, $\Phi_M$ may not belong to a ground state of the entire system. The global constraint for no quasiparticle in the bulk may mix local basis on different boundary components. In the following, without loss of generality, we consider the cases on a disk and a cylinder, respectively, with no quasiparticles in the bulk.
	
	On a disk, we can apply the $\mathcal{T}$ transformation to shrink the bulk graph to a single plaquette, bounded by a circle with outward open edges, as in the equation below. The ground state is non-degenerate and expressed by
	\begin{equation}
	\label{eq:GroundStateDisk}
	\ket{\Phi}=\sum_M\rmd_M \ket{\Phi_M}
	\end{equation}
	\begin{align}
	\label{eq:GroundStateWF}
	\Phi_M\bpm\GroundStateDiskAA\epm
	=
	\rmu_{a_1}\rmu_{a_2}\dots
	[\rho_{M}]^{a_1}_{l_1l_2}[\rho_{M}]^{a_2}_{l_2l_3}\dots
	\end{align}
	
	A cylinder with $A$- and $B$-boundaries can be effectively studied on a circular wall (see Fig. \ref{fig:GroundStateCylinder}). The ground states may be degenerate. The ground state degeneracy (GSD) in terms of $f$ and $G$ is explicitly computed by
	\begin{align}\label{eq:computeGSD}
	&GSD
	=\frac{1}{\rmd_A^2}
	\sum_{st}\rmv_s\rmv_t\sum_{ii'jj'}\rmd_i\rmd_j\rmd_{i'}\rmd_{j'}
	\nonumber\\
	&\quad\times\sum_{aa'bb'}\rmv_a\rmv_b\rmv_{a'}\rmv_{b'}
	f_{s^* {a'}^* a}f_{a' s {a}^*}g_{t^* {b'}^* b}g_{b' t {b}^*}\nonumber\\
	&\quad\times
	G^{j^* i b^*}_{t^* {b'}^* j'}
	G^{i^* a^* j}_{j' {b'}^* i'}
	G^{{{j'}} {{b}} {i'}^*}_{{b'}^* i^* t}
	G^{s^* {a'}^* a}_{j' {i'}^* i}
	G^{s^* i {i'}^*}_{{b}^* {j'}^* j}
	G^{{{a}} {{i}} {j}^*}_{{j'}^* s^* {a'}^*}
	\end{align}
	
	The global constraint on the $Mod_A$ and $Mod_B$ basis leads to a (slightly generalized) $(A,B)$-bimodule structure $P$ satisfying
	\begin{equation}\label{eq:BimoduleDef}
	\Tact\ExpBimoduleDef
	=
	\ExpBimoduleDefAA
	\end{equation}
	for $j_1,j_2\in L$,$a\in L_A$, and $b,c\in L_B$. Note that a box $P$ is not a $4$-valent vertex but a composition of two trivalent vertices (in general, two trivalent vertices may carry extra indices). We denote The collection of all $(A，B）$-bimodules by $Mod_{A|B}$. GSD  equals the total number of $(A,B)$-bimodules. The ground state basis is similarly expressed as in eq. \eqref{eq:GroundStateWF}, with $\rho$ replaced by $P$.
	
	{\bf Boundary Excitations.} The elementary boundary excitations are characterized by topological quasiparticles. On an $A$-boundary component, quasiparticle species are identified with the $(A,A)$-bimodules. We construct a creation operator $W_M$ to create a pair of quasiparticles:
	\begin{equation}\label{eq:creation}
	W_M\ExpQuasiparticleBasis=
	\Tact{\frac{1}{\rmd_A}\ExpQuasiparticleCreation}.
	\end{equation}
	for $M\in Mod_{A|A}$. In this example, the operator $W_M$ creates an $M$-type and $M^*$-type quasiparticles on both neighboring open edges of the middle open edge. By acting creation operators on ground states, we get an elementary boundary excitation basis $W_{M}\ket{\Phi}$. 
	
	Quasiparticles can move along the boundary under the hopping operator $H_M$ defined by
	
	\begin{equation}\label{eq:braiding}
	H_M\ExpQuasiparticleBasisAA=
	\Tact{\ExpQuasiparticleHoppingAA}.
	\end{equation}
	that hops an $M$-type quasiparticle initially at the bottom open edge upward across the edge.
	
	{\bf An example.} Consider the input data for Fibonacci model has Label set $L=\{0,2\}$, sometimes denoted by $\{\mathbf{1},\tau\}$. Let $\phi=\frac{1+\sqrt{5}}{2}$ be the golden ratio. The quantum dimensions are $\rmd_0=1$ and $\rmd_2=\phi$. The fusion rules are $\delta_{000}=\delta_{022}=\delta_{222}=1,\delta_{002}=0$.
	and the nonzero independent $6j$-symbols $G$ are given by
	\begin{align}
	\label{Fib6js}
	G^{000}_{000}=1,
	G^{022}_{022}=G^{022}_{222}=1/\phi,
	\nonumber\\
	G^{000}_{222}=1/\sqrt{\phi},
	G^{222}_{222}=-1/{{\phi}^2}.
	\end{align}
	
	There are two Frobenius algebras:  $A_1=0$, or $A_2=0\oplus 2$. Each $j\in L$ is an $A_1-$module, with action $[\rho_j]^0_{jj}=1$. For $A_2$, $L_A=\{0,2\}$, and $f_{222}=\phi^{-3/4}$. The Frobenius algebra $A_2$ has two modules: (1) $M_0$ is $A_2$ itself, with $[\rho_{M_0}]^a_{jk}=f_{ak^*j}$; (2). $M_1=2$, with $[\rho_{M_1}]^2_{22}=-\phi^{1/4}$. The two Frobenius algebras $A_1$ and $A_2$ give rise to equivalent boundary conditions. They both lead to two boundary quasiparticles species and $GSD=2$ on cylinder. 
	
	{\bf Discussion.} Here we first elaborate on our motivations. Gapping conditions and GSD of Abelian ITOs on open surfaces have recently been understood\cite{Kitaev2012,Levin2013,Kong2013,Barkeshli,Barkeshli2013c,Wang2012,HungWan2013a,Iadecola2014,Barkeshli2014,Barkeshli2014a}. For non-Abelian ITOs on open surfaces, the gapping conditions and GSD counting have recently been solved by the mechanism of anyon  condensation\cite{HungWan2014,HungWan2015a} and by solving certain algebraic equations\cite{Lan2014}. These studies of ITOs on open surfaces are however algebraic and non-dynamical, which limits their applicability, because they lack a Hamiltonian with explicit boundary terms.  
	
	On the other hand, the ground states of an ITO can be effectively described by a continuum Chern-Simons gauge theory. A Chern-Simons theory on an open surface must contain a boundary term  \cite{elitzur1989}, otherwise the bulk Chern-Simons action is not gauge invariant in the presence of spatial boundary. This fact is usually  interpreted as the holographic correspondence between the bulk and the boundary. Such holography exists generally in ITOs in two spatial dimensions. Dynamical theories of topological orders are nevertheless usually formulated using discrete Hamiltonian models. It is thus desirable to demonstrate how the holographic principle works in discrete dynamical models for 2d ITOs. 
	
	Our approach has the following advantages over the existing approaches to gapped ITOs on open surfaces. First, our approach depends on the input data of the model, respecting the usual Hamiltonian dynamics. Second, our boundary Hamiltonians, together with the input data, automatically classifies the gapped boundaries and domain walls of ITOs. Third, by solving the total Hamiltonian, we can obtain the explicit wave functions of the ground and excited states, all in the form of tensor network states. This will provide us a very detailed dynamic understanding of the stationary topological states of the whole bounded system, especially about what is happening on and near the boundary. For example, our model will enable us to study the boundary excitations explicitly. Also anyon condensation will be understood at more microscopic scales. These studies will be reported later separately. Moreover, certain Abelian\cite{Luo2016} and non-Abelian ITOs\cite{Lesanovsky2012,Li2017} on the torus have recently been experimentally simulated on physical systems by imposing periodic boundary conditions. Our approach would make experimental simulation of ITOs on open surfaces possible, and may help construct new quantum computing codes using the boundary states.  
	
	Note added: In preparation of this letter, we noticed a very recent work \cite{Wang2017} by Wang, Wen and Witten that constructed the gapped interfaces between symmetry protected topological and symmetry enriched topological states for finite groups. 
	
	\acknowledgements{We appreciate Zhenghan Wang, Liang Kong, Ren Pankovich, Zhu-Xi Luo, and Ethan Lake for discussions.}

	\bibliographystyle{apsrev}

\end{document}